\documentstyle[aps,multicol,epsfig]{revtex}

\begin{document}

\title{
Intrinsic noise-induced phase transitions: beyond the noise interpretation.}

\author{O. Carrillo$^1$, M. Iba\~nes$^1$, J. Garc\'{\i}a-Ojalvo$^{2,3}$, J.
Casademunt$^1$, J.M. Sancho$^1$}

\address
{$^1$Departament d'Estructura i Constituents de la Mat\`eria,
Universitat de Barcelona, Diagonal 647, E--08028 Barcelona, Spain\\
$^2$Departament de F\'{\i}sica i Enginyeria Nuclear,
Universitat Polit\`ecnica de Catalunya,
Colom 11, E--08222 Terrassa, Spain
$^3$Center for Applied Mathematics, Cornell University, 657 Rhodes
Hall, Ithaca, New York 14853}

\date{\today}

\maketitle

\begin{abstract}
We discuss intrinsic noise effects in stochastic multiplicative-noise partial
differential equations, which are qualitatively independent of the noise
interpretation (It\^o vs. Stratonovich),
in particular in the context of noise-induced ordering phase transitions. We
study a model which, contrary to all cases known so far, exhibits such ordering
transitions when the noise is interpreted not only according to Stratonovich,
but also to It\^o. The main feature of this model is the absence of a linear
instability at the transition point. The dynamical properties of the resulting
noise-induced growth processes are studied and compared in the two
interpretations and with a reference Ginzburg-Landau type model. A detailed
discussion of new numerical algorithms used in both interpretations is also
presented.
\end{abstract}


%\maketitle

\begin{multicols}{2}

\section{Introduction}

An important feature of nonlinear systems is their ability to
sustain an organized behavior even in the presence of a substantial
amount of randomly fluctuating influences. Even more strikingly,
systems which in the absence of fluctuations exhibit a disordered
behavior can experience, under certain conditions, the emergence of
spatiotemporal order upon addition of a suitable amount of noise
\cite{nises}. The most basic manifestation of this fact is the
existence of ordering phase transitions induced by noise in dynamical
systems with spatial degrees of freedom \cite{chris,nio}. These transitions
bring the system from a disordered to an ordered phase as the intensity
of the noise increases, contrary to naive intuition. By disordered
(ordered) phase we mean for example the homogeneous zero (non-zero) state
corresponding to the coarse graining of a spin field with random
(uniform) orientation.

Ordering phase transitions are usually
driven by multiplicative noise terms, which depend on the system's
variables \cite{notapie}. But the stochastic integrals associated with
stochastic
differential equations with multiplicative noise are not uniquely
defined \cite{vankampen}. Among the many interpretations that can be
given to these integrals, two are frequently used: the Stratonovich
interpretation which follows  the standard rules of calculus,
but gives rise to non-intuitive statistical properties of
the noise terms, and  the It\^o interpretation which avoids these
problems, but at the expense of requiring new rules of calculus.

Beyond the technical mathematical definitions, the physical implications of
both noise prescriptions boil down to an important fact. The Stratonovich
prescription for white noise yields the result one would get for a
time-correlated noise in the limit of vanishing correlation time. The key point
is that, as soon as the noise is slightly correlated, the stochastic variables
defined by the corresponding Langevin equation build up correlations with the
noise variable at equal time. This immediately implies that the multiplicative
noise terms  in the equation have a nonzero mean, even with a zero-mean noise.
The result is the so-called Stratonovich drift, a net force induced by noise
which is at the heart of most noise-induced phenomena, in particular concerning
noise-induced ordering transitions.

As for a given stochastic differential equation with multiplicative
noise the results do depend on the
interpretation, a preliminary analysis of the physical problem has to be
performed to make a judicious choice. Our experience indicates that there
are a minimum of three possible situations:
\begin{itemize}
\item If we start with a well established deterministic differential equation
and some controlled parameter is allowed to fluctuate (experimental or
realistic external noise) one would always expect the noise to have a
high--frequency cut-off and as a consequence the Stratonovich
interpretation is usually argued to be the reasonable choice.
\item
If the starting scheme is a master equation which is approximated by a
Fokker-Planck equation, then one can write a stochastic differential
equation with multiplicative noise in the It\^o interpretation.
This happens, for instance, in front
propagation problems
on a lattice \cite{doering1}.
\item
Moreover quite often our initial scheme is a set of stochastic differential
equations, and we would like to simplify the problem eliminating the
most irrelevant fast variables (those with a very short time scale). The
interpretation of the final stochastic differential equation will depend on
the order in which this procedure is performed with respect to the white
noise limit. This is indeed a nontrivial task.
\end{itemize}

Since the Stratonovich drift
can drastically modify the behavior of systems, and since it may not always be
obvious what the appropriate noise prescription is in a given problem, it is
particularly important to distinguish which noise effects are {\em intrinsic},
in the sense of occurring regardless  the noise
interpretation, and which
ones are strictly
associated to the Stratonovich drift. In other words, it is important to
elucidate when the noise interpretation may only affect the quantitative
behavior, and when it may indeed change the problem at a qualitative level.

For the case of noise-induced phase-transitions, the noise prescription used so
far in the literature is that of Stratonovich. Nevertheless, it could be argued
that if the noise has an internal origin, one should in principle expect It\^o
noise too, so it would be good to establish whether, in the latter case,
noise-induced transitions can occur. We will see that this is indeed the case
for a recently discovered class of noise--induced phase
transitions. From a theoretical
point of view, it is also important to deal with It\^o noise since then the
continuum white-noise limit is either well defined or less singular than in the
Stratonovich case \cite{nualart}. This has important consequences in order
to
establish when the macroscopic observables will carry out a nontrivial,
singular dependence on the spatial cut-off of the noise (Stratonovich case)
and
when such residual dependence will be weak (It\^o case) \cite{rocco}.

Few contributions have appeared in the physics literature on It\^o calculus in
extended systems. A comparative discussion about the mathematical problems
involved in the two interpretations appeared in \cite{doering}. The role of the
multiplicative noise in the It\^o interpretation has been analyzed in the
context of spatiotemporal intermittency \cite{palma} and front propagation
\cite{anxo}. Dynamical renormalization group calculations were presented in
\cite{grinstein}. However, noise-induced ordering phase transitions had been
reported so far only in the framework of the Stratonovich interpretation
\cite{nises,chris,nio}. In that case, the mechanism underneath these
transitions is that the multiplicative noise term has a non-zero average value,
which produces a short-time instability of the disordered phase and induces the
ordered phase to arise \cite{chris,nio,kramers}. The instability can be linear
\cite{chris,marta2} or nonlinear \cite{nio,oliver}, but is in any case induced
by the so-called Stratonovich shift. Due to the absence of such a drift, the
It\^o interpretation does not present this type of noise-induced ordered phase,
or any other spatially ordered state \cite{mamunoz}.

Recently, however, a new type of noise-induced phase transition has been found
which does not occur via an instability of the disordered phase
\cite{prlmarta}. Here, the ordered phase arises due to the balance between the
relaxing deterministic forces pushing the system toward the disordered state,
and the activating multiplicative fluctuations pulling the field away from that
state, in a type of entropy-driven phase transition (EDPT). This behavior is
the spatio-temporal extension of noise-induced transitions in purely temporal,
zero-dimensional systems, where the probability distribution of the
time-dependent variable exhibits a change in the number and type of its extrema
as noise intensity varies \cite{horsthemke}. A key idea in the model here
studied is that the bimodality in the stationary probability density is not
associated to a potential barrier, but has a dynamical origin. In fact, the
dependence of the multiplicative noise term on the field is such that, for
sufficiently large noise strength, the system escapes more easily from the
central region than from the sides, despite the fact that the deterministic
force always drives the system towards the center. As a result the peaks of the
probability density are off-center. An important difference with the usual
bimodality associated to a potential barrier is that in our case the
characteristic relaxation time scales for the zero-dimensional model are of
order one (${\cal{O}}(\varepsilon^0)$) as opposed to
${\cal{O}}(\exp(1/\varepsilon)$) which is
characteristic of activation processes, $\varepsilon$ being a
generic measure of the noise strength.
In the spatially extended case, the spatial (diffusive) coupling of
the field introduces an additional crucial ingredient, namely it freezes the
domains impeding the fast relaxation process of the zero-dimensional case. This
gives rise to a well-defined, stable interface which then drives the much
slower domain-growth dynamics.
Since no Stratonovich drift is required to induce this effect (as opposed for
instance to the case of Ref.\cite{nises} where it takes the form of an
effective
barrier) it is to be expected that the corresponding class of model exhibiting
this behavior should also display noise-induced ordering in the It\^o
interpretation. In this paper we show that this is indeed the case, by
comparing the behavior of the model introduced in \cite{prlmarta} for both the
It\^o and Stratonovich interpretations with that of a standard Ginzburg-Landau
model with multiplicative noise (section II). We also analyze in detail the
dynamical properties of the growth processes arising from the noise-induced
ordering transitions in the two cases (section III), which will be shown to
share universal characteristics (i.e. growth exponents) but differ in
non-universal features (such as power-law prefactors). Finally, algorithms that
have been specially developed for generating the results presented in this
paper, for both the Stratonovich and It\^o interpretations, are described in
detail in the Appendix.

\section{Theoretical analysis}

We will use a model of a class of systems for which the steady-state
probability distribution can be obtained {\it exactly}. As a consequence,
the existence of a phase transition in this kind of systems can be studied
without any dynamical reference.

Our model corresponds to a relaxational flow in a free--energy potential
${\cal F}(\{\phi \})$,
with a field-dependent kinetic coefficient $\Gamma(\phi)$ and
a fluctuating term fulfilling a fluctuation-dissipation relation
\cite{prlmarta}. The model is defined by the following
stochastic partial differential equation:
\begin{equation}
\frac{\partial \phi (\vec x,t)}{\partial t} = -
\Gamma(\phi(\vec x,t)) \frac{\delta
\cal{F}}{\delta \phi(\vec x,t)} + \Gamma(\phi(\vec x,t))^{1/2}
\xi(\vec x,t)
\end{equation}
We suppose that the noise $\xi(\vec x,t)$ is Gaussian, with zero mean and
correlation
\begin{equation}
\langle \xi(\vec x,t) \xi(\vec x',t') \rangle = 2 \sigma^{2}
\delta(\vec x - \vec x') \delta(t - t')\,,
\end{equation}
where $\sigma^{2}$ is the noise intensity.
Moreover, we choose the following form for the free-energy potential
$\cal{F}$,
\begin{equation}
{\cal{F}} = \int d^{d}\vec x \left\{ V_{0}(\phi(\vec x,t)) +
\frac{D}{4d}
\left[
\vec{\nabla} \phi (\vec x,t)  \right]^{2}
\right\}\,.
\end{equation}

Since we are dealing with spatially uncorrelated
noise, we perform the analysis in a discrete space in order to avoid
singularities \cite{doering}. In a d--dimensional
square lattice of mesh size $\Delta x$ and $N=L^{d}$ cells, our model
reads
\begin{equation}
\frac{d\, \phi_i}{d\,t} = -\Gamma_i \frac{\partial {F}}{\partial \phi_i} +
\Gamma_i^{1/2}\xi_i(t)\,,
\label{eq:motion}
\end{equation}
where only one index is used to label the cells, $\phi_i\equiv\phi(\vec x_i)$,
$\Gamma_i \equiv \Gamma(\phi_i)$, and the noise satisfies the correlation:
\begin{equation}
\langle \xi_i(t) \xi_j(t') \rangle = 2 \sigma^{2} \frac{\delta_{ij}}
{\Delta x^{d}}\delta(t-t')\,.
\label{eq:noiseNdis}
\end{equation}
In discrete space, the free energy has the form
\begin{equation}
{F}(\{\phi\}) = \sum_{i=1}^{N} \left [V_0(\phi_i) + \frac{D}{4d\Delta x^{2}}
\sum_{j\in nn^{+}(i)}(\phi_j-\phi_i)^{2} \right]\,,
\label{freenergyNd}
\end{equation}
where the gradient term is approximated by the sum over nearest neighbors
on the lattice in a standard way,
$|\vec \nabla\phi|^{2}\to\sum_{j \in
nn^{+}(i)}\frac{(\phi_j-\phi_i)^{2}}{\Delta
x^{2}}$
, and $nn^{+}(i)$ stands for the $d$--nearest neighbors of $i$ in the
positive direction of each axis.
For simplicity, we choose a monostable local potential,
\begin{equation}
V_{0}(\phi) = \frac{a}{2} \phi^{2}\,,
\end{equation}
where $a>0$.
Finally, the kinetic coefficient $\Gamma(\phi)$ is taken to depend on
the field in the following way \cite{prlmarta}
\begin{equation}
\Gamma( \phi ) = \frac{1}{1 + c \phi^{2}}\,.
\label{gamma}
\end{equation}
This functional dependence of the kinetic coefficient favors diffusion
due to fluctuations in the disordered state.

Our objective now is to study the equation (\ref{eq:motion}) in the
Stratonovich and It\^o stochastic interpretations.
The corresponding Fokker--Planck equation for the probability
density of the field $P(\{\phi\},t)$ can be written in a unified notation
for both interpretations \cite{vankampen},
\begin{eqnarray}
\frac{\partial P}{\partial t} = \sum_i \frac{\partial}{\partial \phi_i}
\left[\Gamma_i \frac{\partial {F}}{\partial \phi_i} P +
\frac{B \sigma^{2}}{\Delta x^{d}}\Gamma_i^{1/2}
\frac{\partial \Gamma_i}{\partial \phi_i} P\right.
\nonumber
\\
\left.+ \frac{\sigma^2}{\Delta x^d}
\frac{\partial}{\partial {\phi_i}} \Gamma_i P \right]\,,
\label{probd}
\end{eqnarray}
where $B=1$ for the Stratonovich interpretation and $B=2$ in the It\^o case.

If no probability flux is present, the stationary solution $P_{\rm st}$
of (\ref{probd}) satisfies
\begin{equation}
\left(\frac{\partial {F}}{\partial \phi_i}+ \frac{B\sigma^2}{2\Delta x^d}
\frac{\partial \ln \Gamma_i}{\partial \phi_i} \right) P_{\rm st} +
\frac{\sigma^2}{\Delta x^d} \frac{\partial P_{\rm st}}{\partial \phi_i}=0\,.
\label{Stratoprobd3}
\end{equation}
The solution of this equation is
\begin{equation}
\label{pstNd}
P_{\rm st}(\{\phi\}) \sim {\rm e}^{-{F}_{\rm eff}\Delta x^d/\sigma^2}\,,
\end{equation}
where we have introduced the effective free energy
\begin{equation}
{F}_{\rm eff}(\{\phi\}) \equiv {F}(\{\phi\}) + \frac{B\sigma^2}{2\Delta
x^d} \sum_{i=1}^N \ln \Gamma_i\,.
\label{FeffNd}
\end{equation}

The above expressions can be written in continuum space as
\begin{equation}
P_{\rm st}(\{\phi\}) \sim {\rm e}^{-{\cal F}_{\rm eff}/\sigma^2}\,,
\qquad\qquad
\label{pstNc}
\end{equation}
\begin{equation}
{\cal F^S}_{\rm eff}(\{\phi\}) \equiv {\cal F}(\{\phi\}) +
\frac{B{\sigma_0}^2}{2}
\int \, d^d x \ln \Gamma(\phi(\vec x))\,,
\label{FeffNc}
\end{equation}
where ${\sigma_0}^2\equiv \sigma^2/\Delta x^d$ stands for the effective
noise intensity of a spatially white noise in a discrete space.

We have thus seen that the stationary multivariate probability
distribution can be obtained exactly in both the It\^o and Stratonovich
interpretations
for the spatially extended EDPT model, and that both lead to very
similar qualitative results. The only difference is an extra factor $2$ in
the new term  of the effective potential in the It\^o interpretation. As
is already known \cite{prlmarta}, the EDPT model presents a continuous
ordering noise--induced phase transition in the Stratonovich interpretation.
But according to the results shown above, and as will be shown in the
following Section,
this model also exhibits an {\it ordering} transition in
the {\it It\^o} interpretation, although the location of the critical point
will be different. We should remark here that, as in the case
of the Stratonovich interpretation \cite{prlmarta}, this phase
transition is not due to a short-time instability of the homogeneous
null phase. Indeed, the linear equation for the first statistical moment
$\langle \phi \rangle$ can be computed to be \cite{nises}
\begin{equation}
\frac{\partial \langle \phi \rangle}{\partial t} = -\left[a +
(2-B)\sigma_{0}^2 c\right]
\langle \phi \rangle + \frac{D}{2d} \nabla^2 \langle \phi \rangle\,.
\label{mean}
\end{equation}
For $a>0$, the homogeneous null solution of this equation is stable for all
noise intensities, both for $B=1$ and $B=2$.
Therefore, the mechanism of this phase transition must be different from
the standard one.

\section{Steady-state behavior}

A standard way of determining the existence of a noise-induced phase
transition is by applying a mean-field approximation to the Langevin or
Fokker-Planck equations of the system \cite{nises,chris}.
In the present case, however, since we have obtained the exact multivariate
probability distribution in both interpretations,
we will implement that approximation directly on the effective potential
derived from (\ref{FeffNd}).

\begin{figure}[htb]
\narrowtext
\epsfig{file=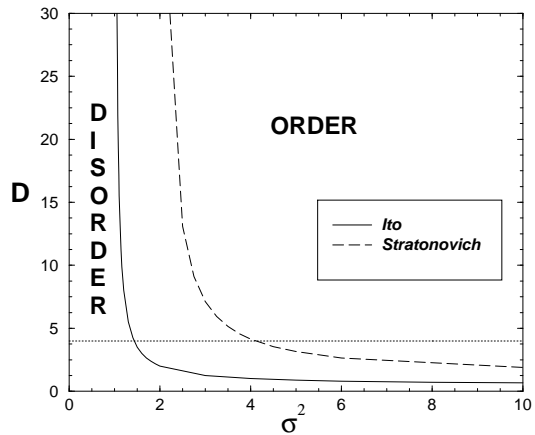,width=6.9cm}
\caption{
Phase diagram of the EDPT model, obtained from a mean-field analysis, in
the It\^o (continuous line) and Stratonovich (dashed line)
interpretations. The horizontal dotted line corresponds to the value
of $D$ used in Fig.~\protect\ref{fig:mfentropy1}.
The parameter values are $a=1$, $c=0.5$, and $\Delta x=1$.}
\label{fig:mfentropy2}
\end{figure}

The mean-field approximation consists in replacing the exact value of
the neighbor field in the Langevin or Fokker-Planck equation
by a common mean-field value $\langle\phi\rangle$.
In the present case, we make such an identification in the neighboring
values of the gradient term appearing in the effective free
energy [see Eqs.~(\ref{freenergyNd}) and (\ref{FeffNd})]:
\begin{equation}
\label{mfappgrad}
\frac{1}{\Delta x^2}\sum_{j\in nn^+(i)}(\phi_j-\phi_i)^2\approx
\frac{2 d}{\Delta x^2}\,(\langle\phi\rangle-\phi_i)^2\,.
\end{equation}
In this way, the effective free energy becomes
\begin{eqnarray}
F_{\rm eff}(\left\{\phi\right\},\langle \phi \rangle) =
\sum_{i=1}^{N}
%\Delta x^{d}
\left\{ V_0(\phi_{i}) +
\frac{{B\sigma_{0}}^{2}}{2}
\ln \Gamma(\phi_{i})
\right.  \nonumber \\
\left. \frac{D}{2 \Delta x^{2}} (\phi_{i} - \langle\phi\rangle)^{2} \right\}
\equiv \sum_{i=1}^{N}
%\Delta x^{d}
V_{\rm eff} (\phi_{i},
\langle\phi\rangle).\quad
\end{eqnarray}
The unknown mean--field value $\langle\phi\rangle$ is obtained
self--consistently, according to
\begin{equation}
\langle\phi\rangle=\int_{-\infty}^{\infty} \phi\,
P_{\rm st}(\phi,\langle\phi\rangle)
\label{self-consNIT}
\end{equation}
where the one-site probability distribution ($P_{\rm st}(\{\phi\})
=\prod_{i=1}^{N}\, P_{\rm st}(\phi_i)$) is given by
\begin{equation}
P_{\rm st}(\phi)\sim {\rm e}^{-V_{\rm eff}/{\sigma_0}^{2}}\,.
\end{equation}
The mean-field predictions for $\langle\phi\rangle$ in the
two-dimensional case are plotted in Fig.~\ref{fig:mfentropy2},
where lines separating the situations where $\langle\phi\rangle=0$
(disorder) and $\langle\phi\rangle\neq 0$ (order) are plotted for
both the It\^o and Stratonovich interpretations in the
space of parameters $D$ and $\sigma^2$. The figure shows
that both interpretations predict a continuous noise--induced
ordering phase transition, which occurs earlier (i.e., for lower noise
intensities) in the It\^o case. In particular, in the large coupling
limit ($D\to\infty$) the transition in the It\^o interpretation takes
place at a critical noise intensity ($\sigma^2_c = a/Bc$, for $\Delta x =
1$),
that is half the critical value in the Stratonovich case, both of which
coincide with the transition point in zero dimensional systems (with
noise intensity $\sigma^2/\Delta x^2$) \cite{horsthemke}.

%%%%%%%%%%%%%%%%%%%%%%%%%%%%%%%%%%%%%%%%%%%%%%%%%%%%%%%%%%%
%%%%%%%%%%%%%%%%%%%%%%%%%%%%%%%%%%%%%%%%%%%%%%%%%%%%%%%%%%%

Note that, in contrast with the usual noise-induced transitions
(which exhibit reentrant phenomena), the transition lines of
Fig. \ref{fig:mfentropy2}
decay monotonously with $\sigma^2$. This implies therefore that no minimum
coupling strength is required in these models for a phase transition to
occur.

In order to validate the results obtained from the mean-field
approximation, we have performed extensive numerical simulations
of model (\ref{eq:motion})-(\ref{gamma}) in both the It\^o and
Stratonovich interpretations. To that end, we have
developed a new type of numerical algorithm suitable for the
implementation of both stochastic interpretations of the multiplicative
noise. The derivation of this algorithm and a comparison with the
well-known Heun algorithm (derived only for the Stratonovich interpretation)
is presented in the Appendix. The simulations have been performed
on a square lattice of $256\times256$ cells of mesh size $\Delta x=1$,
with a time step $\Delta t=0.01$ and periodic boundary conditions
(except when explicitly indicated). Where necessary, we have averaged
over 10 realizations of the noise and the initial random
conditions, corresponding to Gaussian or uniform distributions.
In order to compute the mean field, we first evaluate the spatial
average of the system:
\begin{equation}
\langle\phi(t)\rangle = \frac {1}{N} \left| \sum_{i=1}^N \phi_{i}(t)
\right| \,,
\label{avnum}
\end{equation}
where $N$ is the number of lattice cells, and $\phi_{i}(t)$ is the field
value at the $i$ cell. Once the spatial average reaches a stationary state,
the temporal average is evaluated as
\begin{equation}
\langle\phi\rangle = \frac{1}{T_{M} - T_{m}} \sum^{T_{M}}_{t=T_{m}}
\langle\phi(t)\rangle\,,
\end{equation}
where $T_{M}$ and $T_{m}$ delimit the time interval within the steady-state
regime in which the temporal average is calculated. Afterwards, the
realization average can be computed.
%We make all
%these computations for a given value of the effective multiplicative noise
%intensity, and for a given initial configuration
%of our system.

\begin{figure}[htb]
\narrowtext
\epsfig{file=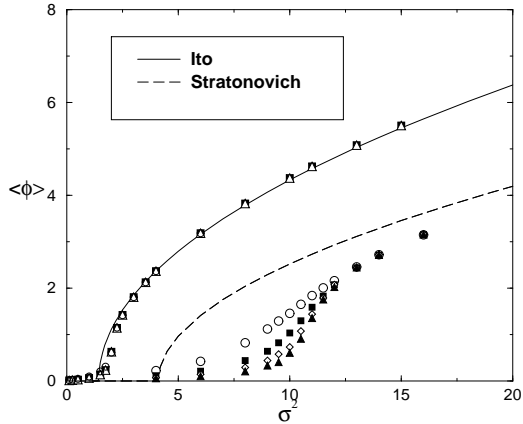,width=7cm}
\caption{
Mean-field and numerical simulation results for the EDPT model in the
It\^o (continuous line) and Stratonovich (dashed line) interpretations.
Simulations have been performed for different system sizes:
$L=16$ (circles), $L=24$ (squares) and $L=32$ (triangles) for It\^o,
and $L=64$ (triangles), $L=48$ (diamonds), $L=32$ (squares) and
$L=16$ (circles) for Stratonovich. $D=4$, and the rest of
parameter values are those of the previous figure.}
\label{fig:mfentropy1}
\end{figure}

The numerical simulation results for the two interpretations are shown in
Fig.~\ref{fig:mfentropy1}, where they are also compared with the predictions
coming from the mean-field approximation. Due to the value of $D$ chosen,
the agreement between the mean-field estimate and the simulations is better
for the It\^o interpretation. In any case, the model exhibits a noise-induced
ordering phase transition for both interpretations, as predicted by
the mean-field approach.

%%%%%%%%%%%%%%%%%%%%%%%%%%%%%%%%%%%%%%%%%%%%%%%%%%%%%%%%%%%
%%%%%%%%%%%%%%%%%%%%%%%%%%%%%%%%%%%%%%%%%%%%%%%%%%%%%%%%%%%

\begin{figure}[htb]
\narrowtext
\centerline{\psfig{figure=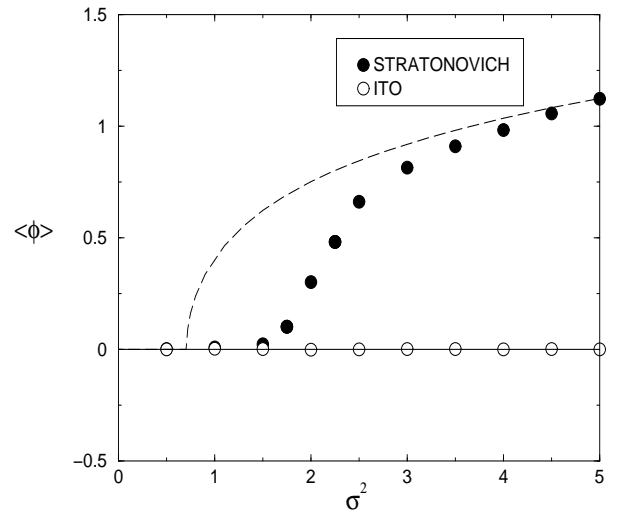,width=8cm,height=7cm}}
\caption{
Bifurcation diagram of the Ginzburg-Landau model in the
It\^o and Stratonovich interpretations. Both mean-field and 2-d simulation
results are shown. The parameters used are $L=30$, $a=-0.2$, $b=1$, $D=4$
and the additive noise intensity $0.5$.
}
\label{4}
\end{figure}

We stress at this point that standard models exhibiting noise-induced
phase transitions caused by short-term instabilities of the disordered
phase do so only in the case of the Stratonovich interpretation. In
order to illustrate this point, we present here for comparison what
happens in the well-known case of the Ginzburg-Landau model with external
multiplicative fluctuations \cite{nio}:
\begin{equation}
\frac{d\phi}{dt} = a\phi - b\phi^{3} + D\nabla^{2}\phi +
\phi\,\xi(\vec x,t) + \eta(\vec x,t)
\label{GLeq}
\end{equation}
where $\eta(\vec x,t)$ and $\xi(\vec x,t)$ are Gaussian and white noises.
This system presents a noise-induced phase transition if we interpret
the noise in the Stratonovich sense, but not if one uses the
It\^o interpretation. This can be seen in Fig.~\ref{4}, where
the two simulations share the same conditions and parameters.
In the It\^o interpretation the ordered parameter $\langle\phi\rangle$
remains always in the
disordered state, due to the fact that the noise-dependent drift
that causes the short-time instability is only present in the
the Stratonovich prescription \cite{mamunoz}.

%%%%%%%%%%%%%%%%%%%%%%%%%%%%%%%%%%%%%%
%%%%%%%%%%%%%%%%%%%%%%%%%%%%%%%%%%%%%%

\section{Domain growth dynamics}

\begin{figure}[htb]
\narrowtext
\begin{center}
{\psfig{figure=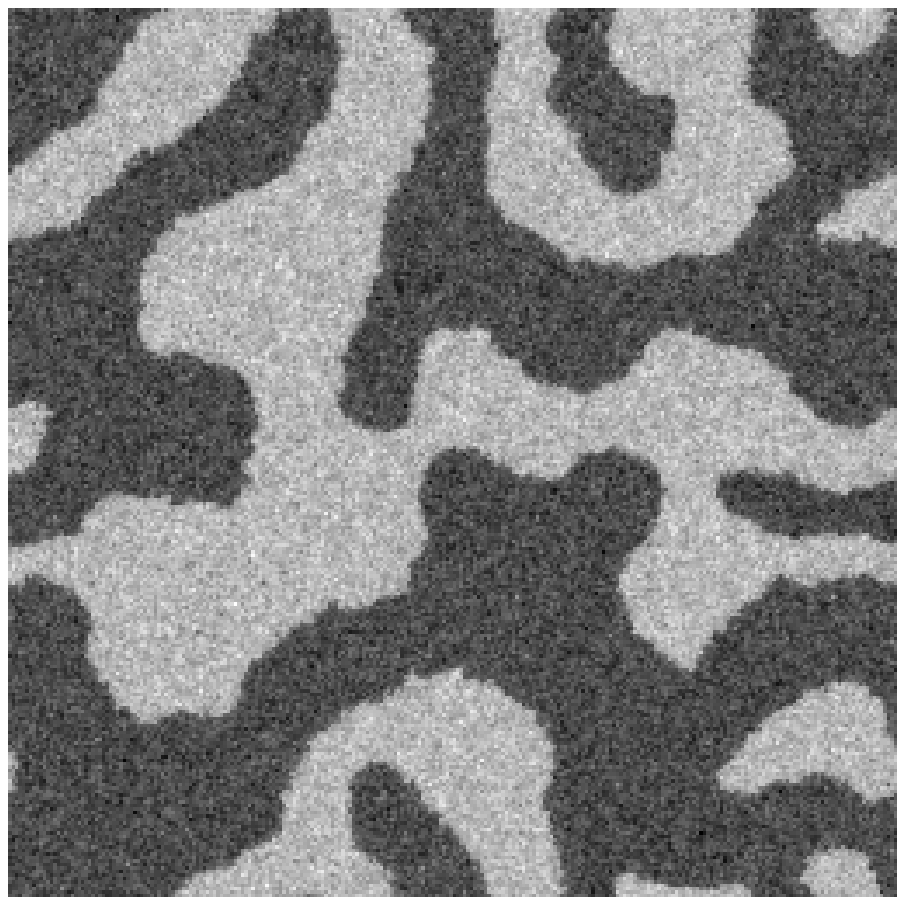,width=4cm,height=4cm}}
\hskip0.001cm
{\psfig{figure=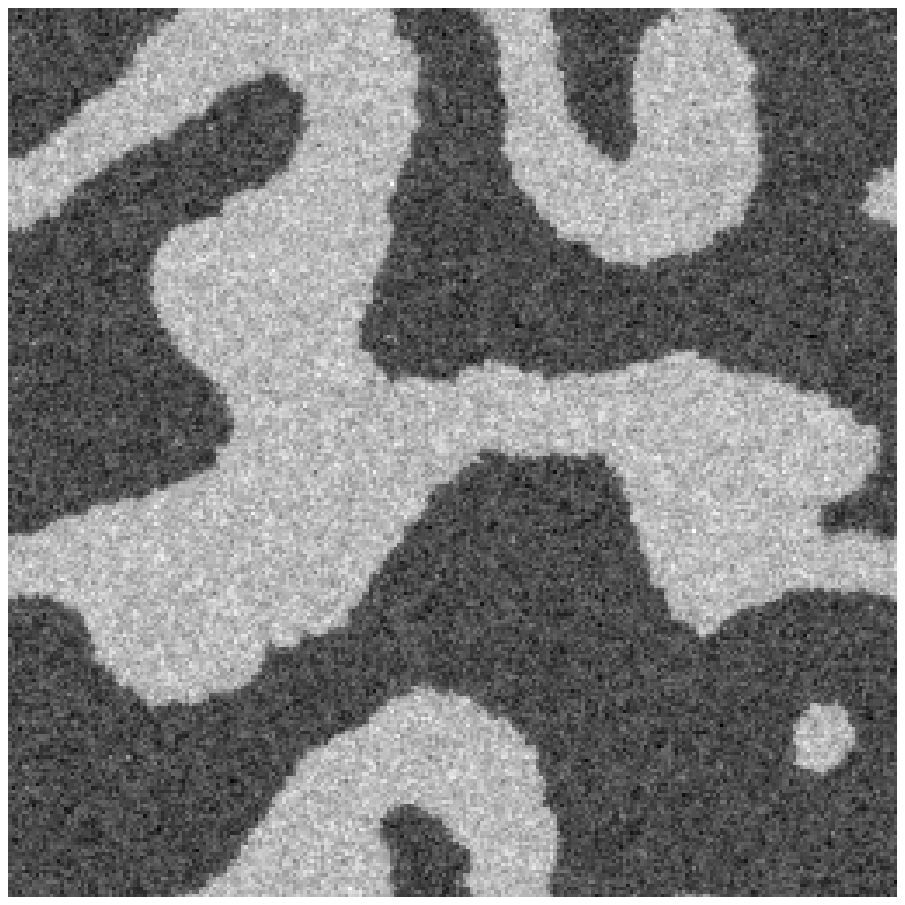,width=4cm,height=4cm}}
\\
{\psfig{figure=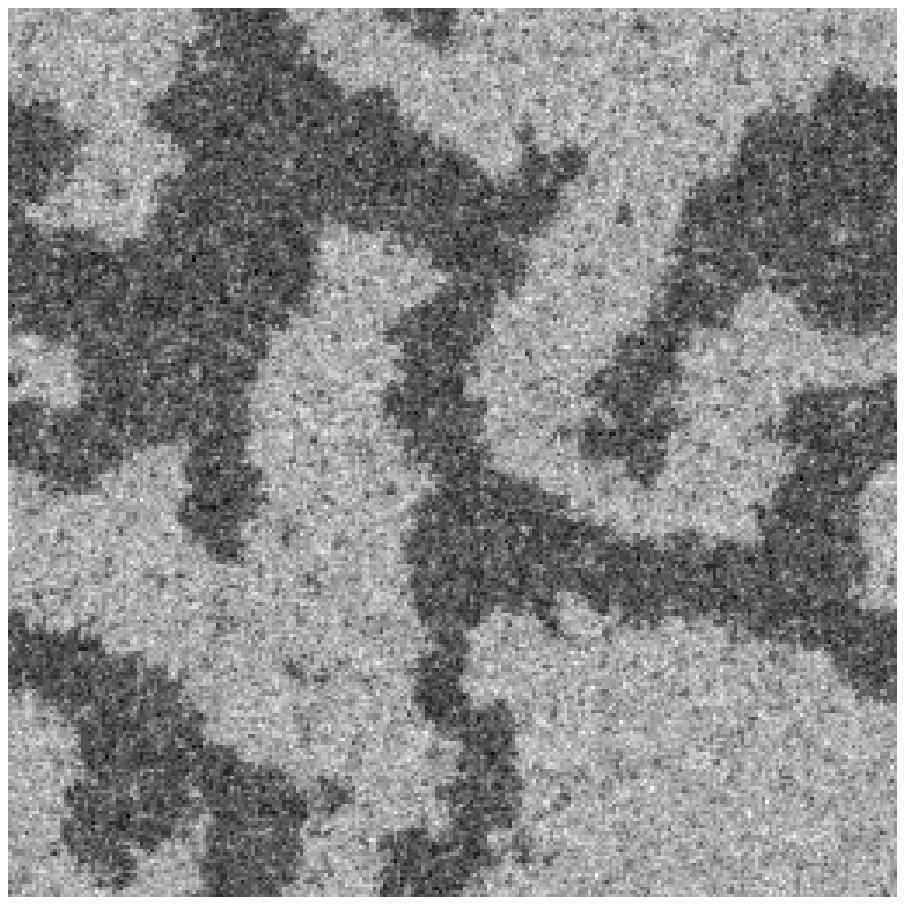,width=4cm,height=4cm}}
\hskip0.001cm
{\psfig{figure=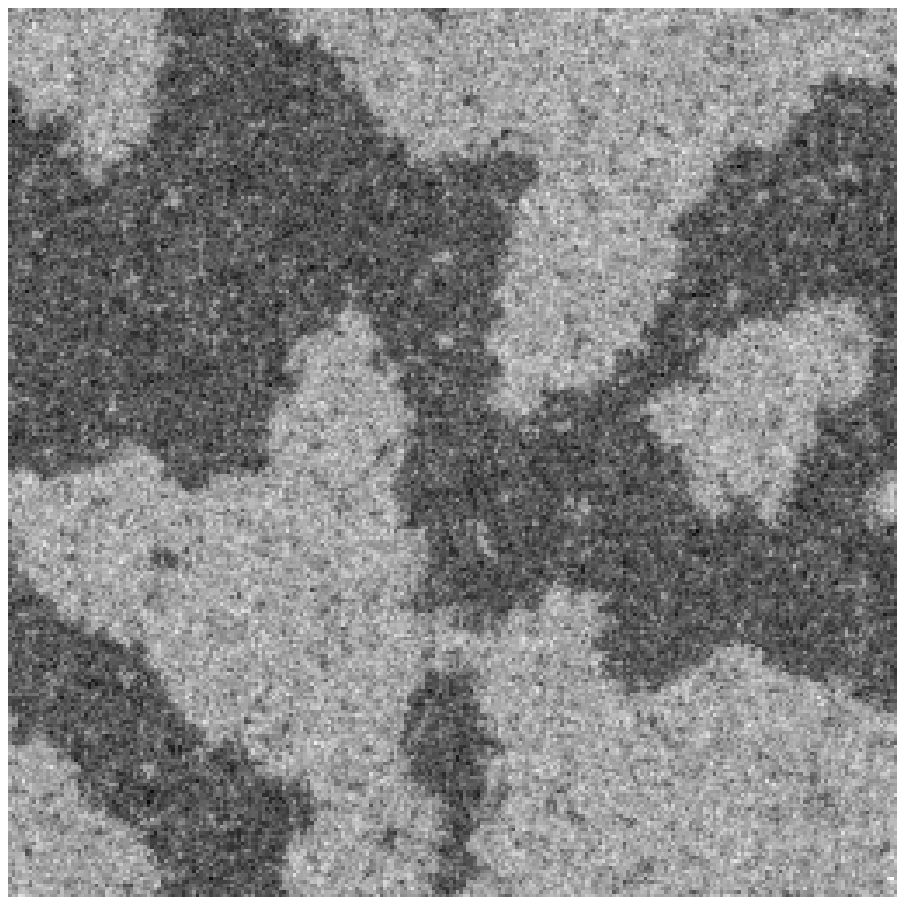,width=4cm,height=4cm}}
\end{center}
\caption{
Snapshots of evolving noise-induced domains for the EDPT model
at $t=750$ (left figures) and $t=1750$ (right figures) in the It\^o
(top) and Stratonovich (bottom) interpretations. Parameters are: $a=1$,
$c=3$, $\sigma^{2}=3.5$ and $L=256$.}
\label{3}
\end{figure}

We have seen that the EDPT model in the presence of external fluctuations
can reach a stationary ordered state described by a non-zero order
parameter $\langle\phi\rangle$, for both the It\^o and Stratonovich
interpretations. This means that, if the system is initially in a
disordered steady state $\langle\phi\rangle=0$
corresponding to a small noise intensity,
%($\sigma^2<a/Bc$)
as the intensity of external fluctuations is increased above its critical
value the system develops
domains of the two new symmetric stationary ordered phases, that grow
with time as shown in Fig.~\ref{3}. The figure shows that the system
behaves differently in the two stochastic interpretations for the same
noise intensity, the It\^o case being much more contrasted due to the
fact that the order parameter is larger than in the Stratonovich case,
which is very noisy.

In this section we are concerned
with the growth of these noise-induced domains.
Although the mechanism that induces the phase transition is different from
those that have been reported before, we can expect that, once the domains
have appeared, their dynamics has the same characteristics as those of the
domain growth following the quench of a system below its order-disorder
transition temperature, as happens in the Ginzburg--Landau model
\cite{marta3}.

%%%%%%%%%%%%%%%%%%%%%%%%%%%%%%%%%%%%%%
%%%%%%%%%%%%%%%%%%%%%%%%%%%%%%%%%%%%%%

For non-conserved order parameter models, one of the domains
grows until it fills the whole system. The mechanism underlying domain
growth in this case is the motion of the interface between domains
caused by the interface structure. The translational velocity of the
domain boundary has been found to be proportional to the mean curvature
of the boundary, and independent of the free energy of the interface.
This can be quantified by the equation of
motion obeyed by the characteristic length (i.e. the average radius) of the
domains of equilibrium phases, $R(t)$, \cite{maxi}
\begin{equation}
\frac{d R}{d t} = A \frac{\Gamma}{R}\,,
\label{CA}
\end{equation}
where $A$ is a model-dependent constant and $\Gamma$ is the
kinetic coefficient multiplying the diffusion term. This expression
leads in a straightforward way to the Allen--Cahn law of domain growth:
\begin{equation}
R(t)\propto \sqrt{2A\Gamma} \,\,t^{1/2}
\label{A-Claw}
\end{equation}
In the time regime where this law is verified, $R(t)$ is the only
characteristic length of the system, and a scaling behavior for its
spatial structure at different times is found. All these results are
known to apply also in the case of standard noise-induced phase
transitions caused by linear instabilities of the homogeneous disordered
phase \cite{marta2}. We want to find out whether the same thing happens
in the EDPT described in this paper.

\begin{figure}[htb]
\narrowtext
\centerline{
\psfig{file=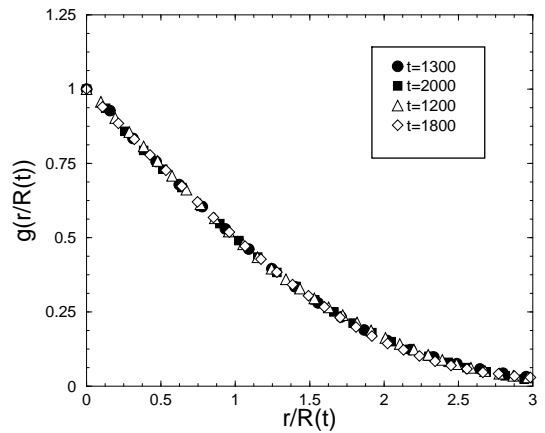,width=7cm}
}
\caption{
Scaled pair correlation function
for the EDPT model in the Stratonovich interpretation
for t=1300 (circles) and t=2000 (squares), and
in the Ito interpretation for
$t=1200$ (triangles) and $t=1800$ (diamonds).
The parameter values are $a=1$, $c=3$, $D=4$, $\sigma^2=3.5$ and
$\Delta x=1$.
}
\label{fig:scalNIT}
\end{figure}

In order to characterize the dynamics of model (\ref{eq:motion}), we
let our system evolve from an initial disordered state, and compute
the isotropic correlation $G(r,t)$ function
at different times. We use the following normalization
\begin{equation}
g(r,t)= \frac{G(r,t)}{G(0,t)}.
\label{normals}
\end{equation}
Let us consider a time regime in which there is only one characteristic
length $R(t)$ in the system, which is related to the average size of
the domains. There are several possible definitions for $R(t)$, but
all of them should lead to the same results. We have chosen $R(t)$ as
the distance at which $g(r,t)$ has half its maximum value. In this time
regime, we can apply the scaling hypothesis for a d-dimensional system
\begin{equation}
g(r,t)=g(r/R(t)),
\label{scaling}
\end{equation}
with no other explicit time dependence. When these relations hold,
the spatial structure of the system at different times is statistically
equivalent, except for a scale factor. Since the domain growth is more
clearly observed far from the critical point, we have taken new parameter
values accordingly. The numerical results in the Stratonovich interpretation
for the scaled pair correlation function
are represented in Fig.~\ref{fig:scalNIT}. As it was shown in
\cite{marta3}, the pair correlation function exhibits a discontinuity in
its first derivative in the presence of noise sources.
We have eliminated this discontinuity by fitting a parabolic function in the
origin ($r=0$, $\Delta x$). The same study has been made in the case
of It\^o interpretation under the same conditions and parameters.

%%%%%%%%%%%%%%%%%%%%%%%%%%%%%%%%%%%%%%%%%%%%%%%%%%%%%%%
\begin{figure}[htb]
\narrowtext
\centerline{\psfig{figure=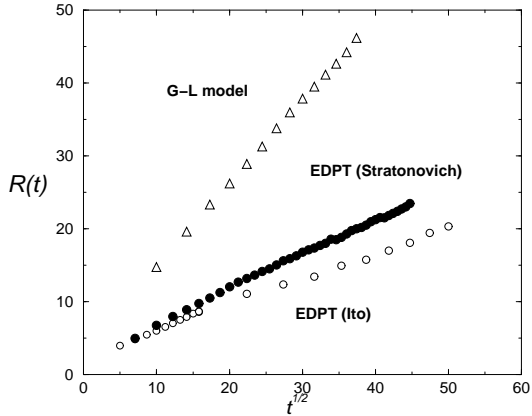,width=7cm}}
\caption{
Allen-Cahn law for the GL model (same parameter values as in Fig.\ref{4}
with
$\epsilon=10^{-4}$ and $\sigma^{2}=0.6$)
and the EDPT model for equal
noise intensities (and different mean fields). The latter is
computed in both the It\^o and Stratonovich interpretations.
The parameter values for the EDPT model are $a = 1$, $c = 3$, $D = 4$,
$\sigma^{2} = 3.5$ and $\Delta x = 1$.
}
\label{ept2}
\end{figure}

%%%%%%%%%%%%%%%%%%%%%%%%%%%%%%%%%%%%%%%%%%%
We now compare the temporal evolution of the characteristic length of the
system $R(t)$ for the two stochastic interpretations. Fig. \ref{ept2}
presents this comparison for equal values of the noise intensity.
From these numerical results we can conclude that the Allen-Cahn law
is satisfied for the two interpretations, and that there is a time regime
in which the system is self-similar. One interesting fact is that
domain evolution in It\^o is slower than in Stratonovich, and in both cases
much slower than the Ginzburg--Landau model. This fact can be
explained looking at the constant prefactor $\sqrt{2A\Gamma}$ of the
Allen-Cahn law (\ref{A-Claw}). In the Ginzburg-Landau model $\Gamma=1$,
but in the EDPT model this quantity is field dependent (\ref{gamma}),
and can be approximated by
\begin{equation}
\Gamma \approx  \frac{1}{1 + c \langle\phi^2\rangle} \approx \frac{1}{1 +
c \langle\phi\rangle^2}.
\label{approx}
\end{equation}
According to this expression, and since for a fixed $\sigma^2$ we have
that
$\langle\phi\rangle_I>\langle\phi\rangle_S$, as a consequence we
should expect the slowest growth for the It\^o EDPT case, and the
fastest one for the Ginzburg-Landau model. This is what we can
see in Fig.~\ref{ept2}.
%%%%%%%%%%%%%%%%%%%%%%%%%%%%%%%%%%%%%%%%%%%
%%%%%%%%%%%%%%%%%%%%%%%%%%%%%%%%%%%%%%%%%%%

\begin{figure}[htb]
\narrowtext
\centerline{\psfig{file=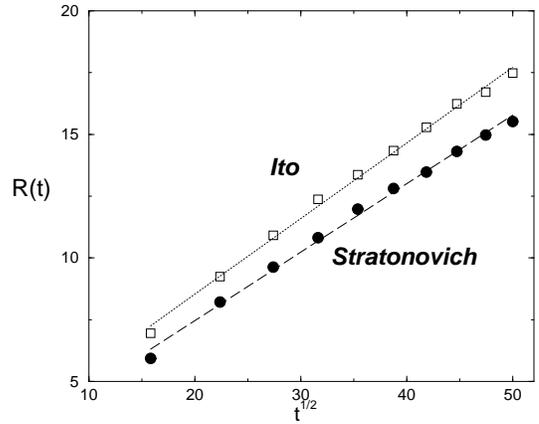,width=7cm}}
\caption{
Allen-Cahn law for the EDPT model for equal mean fields
(and different noise intensities), under both the It\^o and
Stratonovich interpretations.
For Stratonovich $\sigma^{2} = 16$, for It\^o
$\sigma^{2} = 6$.
Other parameter values are: $a=1$, $c=0.5$.}
\label{fig:r}
\end{figure}

In order to eliminate the influence of the stationary mean-field value
$\langle\phi\rangle$
on the growth rate, we have compared the evolution of
the system under the two interpretations using in each case a different
noise intensity, so that the mean field has the same value in the two
cases. The results are shown in Fig \ref{fig:r}, where we have fixed
$\langle\phi\rangle=3.15$ for the two interpretations, for which we need
$\sigma^{2} = 16$ in the Stratonovich interpretation and
$\sigma^{2} = 6$ in the It\^o interpretation.
As can be seen, in both interpretations the system seems to evolve at the
same rhythm, although the slope in It\^o is slightly higher than in
Stratonovich. We think that this small difference is due to the fact
that although  $\langle\phi\rangle$ is the same for both interpretations,
the Stratonovich case is much more fluctuating, and hence we have to expect
a larger $\langle\phi^ 2\rangle$ and accordingly a lower slope.

%%%%%%%%%%%%%%%%%%%%%%%%%%%%%%%%%%%%%%%%%
%%%%%%%%%%%%%%%%%%%%%%%%%%%%%%%%%%%%%%%%%

\section{Comments and conclusions}

\begin{figure}[htb]
\narrowtext
\centerline{\psfig{file=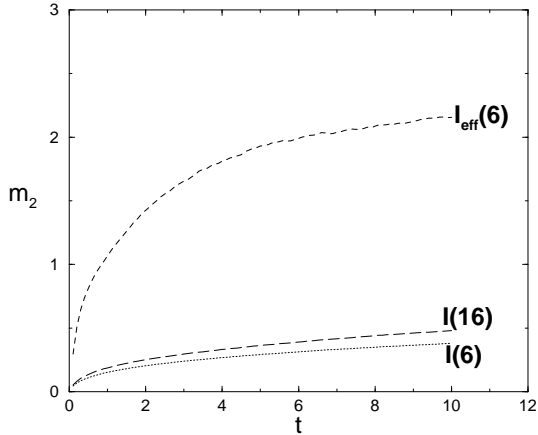,width=7cm}}
\caption{
Transient evolution of the quantity $m_2 = <\phi^2(t)>/<\phi>^2_{st}$
which
measures the emergence of order from homogeneous initial condition
$\phi=0$. The letter $I$ means the It\^o case and the value of the intensity
of
the noise is inside the parenthesis.
}
\label{fig:g0}
\end{figure}

It is worth commenting here that an effective model can be developed
which has the same stationary solutions as the EDPT model, but different
dynamics. The dynamical equation for this effective model with $\Delta x =
1$ is
\begin{equation}
\frac{\partial \phi}{\partial t} = -a \phi + \frac{B \sigma^{2} c \phi}{1 +
c \phi^{2}} + \frac{D}{2d} \nabla^{2} \phi + \xi(\vec x, t)\,,
\end{equation}
where, as before, $B$ is a parameter whose value indicates
the interpretation that we are mimicking ($B=1$ for Stratonovich
and $B=2$ for It\^o). The correlation of noise is given by Eq. (2).
The equation of motion of the mean value of the field in the linear
approximation is
\begin{equation}
\frac{d \langle\phi \rangle}{d t} = (B\sigma^2 c -a)\langle\phi \rangle +
\frac{D}{2d}
\nabla^{2} \langle\phi\rangle\,,
\end{equation}
which tells us that, for $\sigma^2>a/Bc$,
the homogeneous phase $\langle\phi\rangle=0$ is unstable.
This instability does not appear in the EDPT model (see Eq. \ref{mean}).
According to this result, we have to expect an initial transient faster
in this model (as in the Ginzburg-Landau model) than in the EDPT cases.
This fact has been checked numerically and it can be seen in Fig.
\ref{fig:g0}. We can clearly appreciate that the effective model has a
much faster initial transient that the EDPT model, for the same values
of the parameters which is a signature of the different character of the
instability of the initial state. While this observation applies also to
the zero-dimensional version of the model, the crucial ingredient in our
EDPT model is the role of the spatial coupling, which prevents the fast
transition between the two probability peaks. The fact that this
transition occurs in a deterministic time scale in the zero-dimensional case
is what distinguishes the problem from the usual, barrier-crossing
bistability. In the spatially extended case, however, it is precisely
the spatial coupling what generates an effective barrier allowing for
the formation of stable domains, with an interface-driven dynamics.

%%%%%%%%%%%%%%%%%%%%%%%%%%%%%%%%%%%%%%%%%
%%%%%%%%%%%%%%%%%%%%%%%%%%%%%%%%%%%%%%%%%

In conclusion, we have presented a nonequilibrium field model for which one
can compute exactly the stationary probability distribution, and which
exhibits an intrinsic  noise-induced ordering phase transition
irrespective of the
stochastic interpretation of the multiplicative noise term. In particular,
the phase transition is found in the It\^o interpretation, where so far,
noise had only been seen to have a disordering effect. The same model can
be studied changing the diffusive term by the spatial coupling of the
Swift-Hohenberg model in which case a noise-induced pattern transition is
found \cite{buceta}. These type of models do constitute a
generalization of the Horsthemke-Lefever noise induced transitions to
genuine noise induced {\it phase} transitions in extended systems.

\acknowledgments This research was supported by the Direcci\'on General de
Ense\~nanza Superior (Spain) under projects BFM2000-0624, BFM2001-2159,
BXX2000-0638-C02-02 and from the Generalitat de Catalunya (project
2001SGR00223).

\appendix
\section{Stochastic algorithms}

Here we will derive an alternative algorithm which is an extension of the
well-known Heun algorithm, valid for both the It\^o and Stratonovich
interpretations of stochastic differential equations with multiplicative
noise. Our aim is to simulate numerically the following stochastic
differential equation on a d-dimensional lattice,
\begin{equation}
\frac{\partial\phi(\vec x,t)}{\partial t}=f(\phi(\vec x,t),\nabla)
+g(\phi(\vec x,t)) \xi(\vec x,t).
\end{equation}
First of all, we write this equation in a discrete space as follows,
\begin{equation}
\frac{d \phi_{i}(t)}{dt} = f_{i} (\phi(t)) + g_{i}(\phi
(t)) \xi_{i} (t)\,,
\label{sde}
\end{equation}
where $i$ stands for the position inside the lattice, and the noise
correlation is given by Eq. (\ref{eq:noiseNdis}).

The first step in the derivation of the algorithm is to
integrate formally Eq. (\ref{sde}) to get
\begin{eqnarray}
\phi_{i} (t + \Delta t) = \phi_{i} (t) &+&
 \int_{t}^{t + \Delta t} f_i(\phi(t')) dt'
\nonumber \\
&+& \int_{t}^{t + \Delta t} g(\phi(t'))
\xi(t') dt'
\label{eq1}
\end{eqnarray}
The first integral in (\ref{eq1}) is evaluated according to a second-order
predictor-corrector algorithm.
\begin{eqnarray}
\phi_{i} (t + \Delta t) = \phi_{i} (t) &+& \frac{f_{i}(\phi(t)) +
f_{i}(\tilde{\phi}(t))}{2} \Delta t
\nonumber \\
&+& \int_{t}^{t + \Delta t} g(\phi(t'))
\xi(t') dt'
\end{eqnarray}
where $\tilde{\phi}(t)$ is the predictor term defined as the first-order
solution of (\ref{eq1}),
\begin{equation}
\tilde{\phi}_{i} (t) = \phi_{i} (t) + f_{i}(\phi_{i} (t)) \Delta t + g_{i}
(\phi (t)) X_{i}
\label{eqpredictor}
\end{equation}
This expression defines the first equation of the algorithm, which is
independent of the stochastic interpretation. $X_{i}(t)$ is the
Wiener process, defined as
\begin{equation}
X_{i}(t) = \int_{t}^{t + \Delta t} \xi_{i} (t') dt'
\end{equation}
and whose numerical implementation is
\begin{equation}
X_{i}(t)= \sqrt{ \frac{2 \sigma^2 \Delta t}{\Delta x^2}} \alpha_i\,,
\end{equation}
where $\alpha_i$ are independent Gaussian random numbers of zero mean and
unity variance, and they are implemented using \cite{raul}.

The second integral in Eq. (\ref{eq1}) is not well defined, and
one needs to make a prescription for its evaluation, at least up to first
order in $\Delta t$.

The standard Heun algorithm works for the Stratonovich interpretation,
and makes the following assumption
\begin{equation}
\int_{t}^{t+\Delta t} g(\phi(t')) \xi(t') dt'=\left(\frac{g(\phi_{i}(t))
+g(\tilde{\phi}_{i}(t))}{2}\right) X_{i}(t)
\end{equation}
Accordingly, the second equation of this algorithm is
\begin{eqnarray}
\phi_{i}(t + \Delta t) = \phi_{i}(t) +
\frac{f_{i}(\phi (t)) + f_{i}(\tilde{\phi} (t))}{2} \Delta t
\nonumber \\
+ \left(\frac{
g(\phi_{i}(t)) + g(\tilde{\phi}_{i}(t))}{2}\right) X_{i}(t)
\label{eqHeun}
\end{eqnarray}
On the other hand, in the
Stratonovich calculus this integral is interpreted as \cite{vankampen}
\begin{equation}
\int_{t}^{t + \Delta t} g (\phi (t')) \xi(t') dt' = g \left(\frac{
\phi_{i}(t) + \tilde{\phi}_{i}(t)}{2}\right) X_{i}(t)\,,
\end{equation}
so that the second equation of the algorithm is
\begin{eqnarray}
\phi_{i} (t + \Delta t) = \phi_{i}(t) +
\frac{f_{i}(\phi (t)) + f_{i}(\tilde{\phi} (t))}{2} \Delta t
\nonumber \\
+ g \left(\frac{
\phi_{i}(t) + \tilde{\phi}_{i}(t)}{2}\right) X_{i} (t),
\label{eqS}
\end{eqnarray}
which is not exactly the standard Heun algorithm. This is the algorithm
that has been used in this paper for the Stratonovich interpretation
results. Its advantage is that, in contrast to Heun algorithm, our
method has an analogue in the Ito interpretation, for which the same
integral is defined as \cite{vankampen}
\begin{equation}
\int_{t}^{t + \Delta t} g_{i}(\phi (t')) \xi_{i} (t') dt' = g_{i}( \phi
(t) ) X_{i}(t).
\end{equation}
Therefore, the second equation of the algorithm in Ito interpretation
reads
\begin{eqnarray}
\phi_{i}(t + \Delta t) = \phi_{i}(t) + \frac{f_{i}(\phi(t)) +
f_{i}(\tilde{\phi}(t))}{2} \Delta t
\nonumber \\
+ g_{i}( \phi(t) ) X_{i}(t).
\label{eqI}
\end{eqnarray}
Given these results, the algorithm proceeds by evaluating first the
predictor contribution (\ref{eqpredictor}) and, using this value,
computing the corrector term (\ref{eqHeun}), (\ref{eqS}) or
(\ref{eqI}), corresponding to the Heun, Stratonovich or Ito algorithms
respectively. All these three different algorithms are approximations
up to the same order (second order in the deterministic part but first
order in the stochastic one), when properly expanded in powers of
$\Delta t$. One can check that there are no differences,
up to these orders, between the Heun and Stratonovich algorithms, as it
should be. Nevertheless, the Stratonovich prescription has an extra term,
$1/2 g(\phi_i)g'(\phi_i)X_i(t)^2$, with respect the It\^o one, which is of
order $\Delta t$. Our Ito algorithm also agrees with the one presented in
\cite{rao} up to order $\Delta t^{2}$.

\end{multicols}
\end{document}